\input harvmac
\newcount\yearltd\yearltd=\year\advance\yearltd by 0

%\draftmode

% Figure definitions

\input epsf

\newcount\figno
\figno=0
\def\fig#1#2#3{
\par\begingroup\parindent=0pt\leftskip=1cm\rightskip=1cm\parindent=0pt
\baselineskip=11pt
\global\advance\figno by 1
\midinsert
\epsfxsize=#3
\centerline{\epsfbox{#2}}
\vskip 12pt
{\bf Figure \the\figno:} #1\par
\endinsert\endgroup\par
}
\def\figlabel#1{\xdef#1{\the\figno}}

% Other definitions

\noblackbox
\def\IZ{\relax\ifmmode\mathchoice
{\hbox{\cmss Z\kern-.4em Z}}{\hbox{\cmss Z\kern-.4em Z}}
{\lower.9pt\hbox{\cmsss Z\kern-.4em Z}} {\lower1.2pt\hbox{\cmsss
Z\kern-.4em Z}}\else{\cmss Z\kern-.4em Z}\fi}

\font\cmss=cmss10 \font\cmsss=cmss10 at 7pt
\def\IR{\relax{\rm I\kern-.18em R}}

\def\frac#1#2{{#1 \over #2}}

%#\draftmode  
%#\def\journal#1&#2(#3){\unskip, \sl #1\ \bf #2 \rm(19#3) }  
%#\def\andjournal#1&#2(#3){\sl #1~\bf #2 \rm (19#3) }  
%#\def\nextline{\hfil\break}  
%#  

\def\frac#1#2{{#1\over#2}}

%  
%%%%%%%%%%%%%%%%%%%%%%%%%%%%%%%%%%%%  
%  

%\catcode`\@=11  
%\def\slash#1{\mathord{\mathpalette\c@ncel{#1}}}  
\overfullrule=0pt

\def\MM{{\cal M}}  
  
\def\OO{{\cal O}}

\def\SS{{\cal S}}

\def\underrel#1\over#2{\mathrel{\mathop{\kern\z@#1}\limits_{#2}}}

\def\cd{coordinate}

%\catcode`\@=12  
  
%%%%%%%%%%%%%%%%%%%%%%%%%%%%%%%%%%%%%%%%%%%%%%%%%%%%%%%%%%%%%%  
%\def\sdtimes{\mathbin{\hbox{\hskip2pt\vrule height 4.1pt depth -.3pt width  
%.25pt \hskip-2pt$\times$}}}  
%  

\noblackbox

%\AharonyPA
\lref\AharonyPA{ O.~Aharony, M.~Berkooz and E.~Silverstein,
``Multiple-trace operators and non-local string theories,''
JHEP {\bf 0108}, 006 (2001),hep-th/0105309.
%%CITATION = HEP-TH 0105309;%%
}

%\KutasovXU
\lref\KutasovXU{
D.~Kutasov and N.~Seiberg,
``More comments on string theory on $AdS_3$,''
JHEP {\bf 9904}, 008 (1999),hep-th/9903219.
%%CITATION = HEP-TH 9903219;%%
}

%\deBoerPP
\lref\deBoerPP{
J.~de Boer, H.~Ooguri, H.~Robins and J.~Tannenhauser,
``String theory on $AdS_3$,''
JHEP {\bf 9812}, 026 (1998)
,hep-th/9812046.
%%CITATION = HEP-TH 9812046;%%
}

%\KachruYS
\lref\KachruYS{
S.~Kachru and E.~Silverstein,
``4d conformal theories and strings on orbifolds,''
Phys.\ Rev.\ Lett.\  {\bf 80}, 4855 (1998)
,hep-th/9802183.
%%CITATION = HEP-TH 9802183;%%
}

%\GiveonNS
\lref\GiveonNS{
A.~Giveon, D.~Kutasov and N.~Seiberg,
``Comments on string theory on $AdS_3$,''
Adv.\ Theor.\ Math.\ Phys.\  {\bf 2}, 733 (1998)
,hep-th/9806194.
%%CITATION = HEP-TH 9806194;%%
}

\lref\usnext{O. Aharony, M. Berkooz, E. Silverstein, ... work in
progress.}

%\Maldacena
\lref\MaldacenaRE{
J.~Maldacena,
``The large $N$ limit of superconformal field theories and supergravity,''
Adv.\ Theor.\ Math.\ Phys.\  {\bf 2}, 231 (1998)
[Int.\ J.\ Theor.\ Phys.\  {\bf 38}, 1113 (1998)]
,hep-th/9711200.
%%CITATION = HEP-TH 9711200;%%
}

%\DijkgraafJT
\lref\DijkgraafJT{
R.~Dijkgraaf, E.~Verlinde and H.~Verlinde,
``On Moduli Spaces Of Conformal Field Theories With c $\geq$ 1,''
{\it in *Copenhagen 1987, proceedings, perspectives in string theory*,
117-137. }
}

%\BalasubramanianNH
\lref\BalasubramanianNH{
V.~Balasubramanian, M.~Berkooz, A.~Naqvi and M.~J.~Strassler,
``Giant gravitons in conformal field theory,''
,hep-th/0107119.
%%CITATION = HEP-TH 0107119;%%
}

%\LarsenUK
\lref\LarsenUK{
F.~Larsen and E.~J.~Martinec,
``$U(1)$ charges and moduli in the D1-D5 system,''
JHEP {\bf 9906}, 019 (1999)
,hep-th/9905064.
%%CITATION = HEP-TH 9905064;%%
}

%\GiveonUQ
\lref\GiveonUQ{
A.~Giveon, D.~Kutasov and A.~Schwimmer,
``Comments on D-branes in $AdS_3$,''
Nucl.\ Phys.\ B {\bf 615}, 133 (2001)
,hep-th/0106005.
%%CITATION = HEP-TH 0106005;%%
}

%\WittenQJ
\lref\WittenQJ{
E.~Witten,
``Anti-de Sitter space and holography,''
Adv.\ Theor.\ Math.\ Phys.\  {\bf 2}, 253 (1998)
,hep-th/9802150.
%%CITATION = HEP-TH 9802150;%%
}

%\GubserBC
\lref\GubserBC{
S.~S.~Gubser, I.~R.~Klebanov and A.~M.~Polyakov,
``Gauge theory correlators from non-critical string theory,''
Phys.\ Lett.\ B {\bf 428}, 105 (1998)
,hep-th/9802109.
%%CITATION = HEP-TH 9802109;%%
}

\lref\MaldacenaIM{
J. Maldacena, 
``Wilson Loops in Large N Field Theory'',
Phys. Rev. Lett. {\bf 80}, 4859 (1998), 
,hep-th/9803002.
}

\lref\ReyIK{
S. J. Rey and J. Yee, ``Macroscopic Strings as Heavy Quarks in Large N
Gauge Theory and Anti-de-Sitter Supergravity'',
Eur. Phys. J. C {\bf 22}, 379 (2001),
,hep-th/9803001.
} 

\lref\FischlerTB{
W. Fischler and L. Susskind,
``Dilaton Tadpoles, String Condensates, and Scale Invariance 2.''
Phys. Lett. B {\bf 173}, 262, (1986)
}

\lref\FischlerCI{
W. Fischler and L. Susskind,
``Dilaton Tadpoles, String Condensates, and Scale Invariance.''
Phys. Lett. B {\bf 171}, 383, (1986)
}

%\BalasubramanianRT
\lref\BalasubramanianRT{
V.~Balasubramanian, J.~de Boer, E.~Keski-Vakkuri and S.~F.~Ross,
``Supersymmetric conical defects: Towards a string theoretic description
of black hole formation,''
Phys.\ Rev.\ D {\bf 64}, 064011 (2001)
,hep-th/0011217.
%%CITATION = HEP-TH 0011217;%%
}

%\WittenHF
\lref\WittenHF{
E.~Witten,
``Quantum Field Theory And The Jones Polynomial,''
Commun.\ Math.\ Phys.\  {\bf 121}, 351 (1989).
%%CITATION = CMPHA,121,351;%%
}

%\ElitzurNR
\lref\ElitzurNR{
S.~Elitzur, G.~W.~Moore, A.~Schwimmer and N.~Seiberg,
``Remarks On The Canonical Quantization Of The Chern-Simons-Witten Theory,''
Nucl.\ Phys.\ B {\bf 326}, 108 (1989).
%%CITATION = NUPHA,B326,108;%%
}

%\MaldacenaSS
\lref\MaldacenaSS{
J.~Maldacena, G.~W.~Moore and N.~Seiberg,
``D-brane charges in five-brane backgrounds,''
JHEP {\bf 0110}, 005 (2001)
,hep-th/0108152.
%%CITATION = HEP-TH 0108152;%%
}

%\GiveonUQ
\lref\GiveonUQ{
A.~Giveon, D.~Kutasov and A.~Schwimmer,
``Comments on D-branes in $AdS_3$,''
Nucl.\ Phys.\ B {\bf 615}, 133 (2001)
,hep-th/0106005.
%%CITATION = HEP-TH 0106005;%%
}

%\BachasFR
\lref\BachasFR{
C.~Bachas and M.~Petropoulos,
``Anti-de-Sitter D-branes,''
JHEP {\bf 0102}, 025 (2001)
,hep-th/0012234.
%%CITATION = HEP-TH 0012234;%%
}

%\RajaramanEW
\lref\RajaramanEW{
A.~Rajaraman,
``New $AdS_3$ branes and boundary states,''
arXiv:hep-th/0109200.
%%CITATION = HEP-TH 0109200;%%
}

%\ParnachevGW
\lref\ParnachevGW{
A.~Parnachev and D.~A.~Sahakyan,
``Some remarks on D-branes in $AdS_3$,''
JHEP {\bf 0110}, 022 (2001)
,hep-th/0109150.
%%CITATION = HEP-TH 0109150;%%
}

%\RajaramanCR
\lref\RajaramanCR{
A.~Rajaraman and M.~Rozali,
``Boundary states for D-branes in $AdS_3$,''
arXiv:hep-th/0108001.
%%CITATION = HEP-TH 0108001;%%
}

%\HikidaYI
\lref\HikidaYI{
Y.~Hikida and Y.~Sugawara,
``Boundary states of D-branes in $AdS_3$ based on discrete series,''
arXiv:hep-th/0107189.
%%CITATION = HEP-TH 0107189;%%
}

%\LeeXE
\lref\LeeXE{
P.~Lee, H.~Ooguri, J.~W.~Park and J.~Tannenhauser,
``Open strings on $AdS_2$ branes,''
Nucl.\ Phys.\ B {\bf 610}, 3 (2001)
,hep-th/0106129.
%%CITATION = HEP-TH 0106129;%%
}

%\PetropoulosQU
\lref\PetropoulosQU{
P.~M.~Petropoulos and S.~Ribault,
``Some remarks on anti-de Sitter D-branes,''
JHEP {\bf 0107}, 036 (2001)
,hep-th/0105252.
%%CITATION = HEP-TH 0105252;%%
}

%\StanciuNX
\lref\StanciuNX{
S.~Stanciu,
``D-branes in an $AdS_3$ background,''
JHEP {\bf 9909}, 028 (1999)
,hep-th/9901122.
%%CITATION = HEP-TH 9901122;%%
}

%\Figueroa-O'FarrillEI
\lref\FigueroaOFarrillEI{
J.~M.~Figueroa-O'Farrill and S.~Stanciu,
``D-branes in $AdS_3\times S^3\times S^3\times S^1$,''
JHEP {\bf 0004}, 005 (2000)
,hep-th/0001199.
%%CITATION = HEP-TH 0001199;%%
}

%\RyangPX
\lref\RyangPX{
S.~J.~Ryang,
``Nonstatic $AdS_2$ branes and the isometry group of $AdS_3$ spacetime,''
arXiv:hep-th/0110008.
%%CITATION = HEP-TH 0110008;%%
}

%\MaldacenaKY
\lref\MaldacenaKY{
J.~Maldacena, G.~W.~Moore and N.~Seiberg,
``Geometrical interpretation of D-branes in gauged WZW models,''
JHEP {\bf 0107}, 046 (2001)
,hep-th/0105038.
%%CITATION = HEP-TH 0105038;%%
}

%\GiddingsCX
\lref\GiddingsCX{
S.~B.~Giddings and A.~Strominger,
``Loss Of Incoherence And Determination
Of Coupling Constants In Quantum Gravity,''
Nucl.\ Phys.\ B {\bf 307}, 854 (1988).
%%CITATION = NUPHA,B307,854;%%
}

%\GiddingsWV
\lref\GiddingsWV{
S.~B.~Giddings and A.~Strominger,
``Baby Universes, Third Quantization And The Cosmological Constant,''
Nucl.\ Phys.\ B {\bf 321}, 481 (1989).
%%CITATION = NUPHA,B321,481;%%
}

%\ColemanTJ
\lref\ColemanTJ{
S.~R.~Coleman,
``Why There Is Nothing Rather Than Something:
A Theory Of The Cosmological Constant,''
Nucl.\ Phys.\ B {\bf 310}, 643 (1988).
%%CITATION = NUPHA,B310,643;%%
}

%\HawkingMZ
\lref\HawkingMZ{
S.~W.~Hawking,
``Quantum Coherence Down The Wormhole,''
Phys.\ Lett.\ B {\bf 195}, 337 (1987).
%%CITATION = PHLTA,B195,337;%%
}

%\MaldacenaKM
\lref\MaldacenaKM{
J.~Maldacena and H.~Ooguri,
``Strings in $AdS_3$ and the $SL(2,\IR)$ WZW
model. III: Correlation  functions,''
arXiv:hep-th/0111180.
%%CITATION = HEP-TH 0111180;%%
}

\lref\witten{E.~Witten,``Anti-de Sitter space and holography,''
Adv.\ Theor.\ Math.\ Phys.\  {\bf 2} (1998) 253 ,hep-th/9802150.}

\lref\ma{J.~Maldacena,
``The large $N$ limit of superconformal field theories and supergravity,''
Adv.\ Theor.\ Math.\ Phys.\  {\bf 2} (1998) 231
[Int.\ J.\ Theor.\ Phys.\  {\bf 38} (1998) 1113]
,hep-th/9711200.}

\lref\kw{I.~R.~Klebanov and E.~Witten,
``AdS/CFT correspondence and symmetry breaking,''
Nucl.\ Phys.\ B {\bf 556} (1999) 89
,hep-th/9905104.}

\lref\abs{O.~Aharony, M.~Berkooz and E.~Silverstein,
`Multiple-trace operators and non-local string theories,''
JHEP {\bf 0108} (2001) 006
,hep-th/0105309.}

\lref\gkp{S.~S.~Gubser, I.~R.~Klebanov and A.~M.~Polyakov,
``Gauge theory correlators from non-critical string theory,''
Phys.\ Lett.\ B {\bf 428} (1998) 105
,hep-th/9802109.}

%\AharonyTI
\lref\AharonyTI{
O.~Aharony, S.~S.~Gubser, J.~Maldacena, H.~Ooguri and Y.~Oz,
``Large $N$ field theories, string theory and gravity,''
Phys.\ Rept.\  {\bf 323}, 183 (2000)
,hep-th/9905111.
%%CITATION = HEP-TH 9905111;%%
}

\lref\nlst{
O. Aharony, M. Berkooz and E. Silverstein, ``Non-local String Theories
in $AdS_3 \times S^3$ and Stable Non-Supersymmetric Backgrounds'',
,hep-th/0112178}

\lref\evaallan{
A. Adams and E. Silverstein,
``Closed String TAchyons, AdS/CFT and Large N QCD'',
Phys.\ Rev.\ D64:086001, 2001, ,hep-th/0103220
}

%\lref\bfs{M. Bianchi, D.Z. Freedman and K. Skenderis,
%``Holographic Renormalization'',
%hep-th/0112119}

\lref\futwork{O. Aharony, M. Berkooz and E. Silverstein, work in progress}

\lref\bd{
N.D. Birrell and P.C.W. Davies,
``Quantum Fields in Curved Space'',
Cambridge Monographs on Mathematical Physics, 1982}

\lref\witbou{E.~Witten, ``Multi-trace operators, boundary conditions,
 and AdS/CFT correspondence,''  hep-th/0112258.}

\lref\minces{P.~Minces and V.~O.~Rivelles, ``Energy and the AdS/CFT 
correspondence,''
 JHEP {\bf 0112} (2001) 010, hep-th/0110189.}

%\draftmode

%%%%%%%%%%%%%%%%%%%%%%%%%%%%%%%%%%%%%%%%%%%%%%%%%%%%%%%%%%%%%%%%%%%%%%%%%%

\def\myTitle#1#2{\nopagenumbers\abstractfont\hsize=\hstitle\rightline{#1}%
\vskip 0.5in\centerline{\titlefont #2}\abstractfont\vskip .5in\pageno=0}

\myTitle{\vbox{\baselineskip12pt\hbox{hep-th/0112264}
\hbox{RI-12-01}
\hbox{WIS/01/02-JAN-DPP}
}} {\vbox{
        \centerline{``Double-trace'' Deformations, Boundary Conditions}
	\medskip
        \centerline{and Spacetime Singularities}}}
%\medskip
\centerline{Micha Berkooz$^{a,}$\foot{E-mail :
{\tt Micha.Berkooz@weizmann.ac.il.}
Incumbent of the Recanati
career development chair of energy research.}, 
Amit Sever$^{b,}$\foot{E-mail : {\tt asever@cc.huji.ac.il .}} and 
Assaf Shomer$^{b,}$\foot{E-mail : {\tt shomer@cc.huji.ac.il.}}}
\medskip
\centerline{$^{a}$Department of Particle Physics, The Weizmann
Institute of Science, Rehovot 76100, Israel}
\medskip
\centerline{$^{b}$Racah Institute of Physics, The Hebrew University, 
Jerusalem 91904, Israel}

\bigskip
%\bigskip
\noindent

Double-trace deformations of the AdS/CFT duality result in a new
perturbation expansion for string theory, based on a non-local
worldsheet. We discuss some aspects of the deformation in the low
energy gravity approximation, where it appears as a change in the
boundary condition of fields. We relate unique features of the
boundary of $AdS$ to the worldsheet becoming non-local, and conjecture
that non-local worldsheet actions may be generic in other classes of
backgrounds.

\Date{January 2002}

\baselineskip=16pt

\newsec{Introduction}

Non-local string theories \refs{\abs,\nlst} (NLST's) are new string
theories with the following properties
\item{1.} They are described by a non-local worldsheet action.
\item{2.} The familiar string perturbation theory is modified. In addition 
to the usual topologies there are now singular topologies, weighted with a
new expansion parameter. This new parameter is not a modulus.
\item{3.} The theories are maximally non-local on parts of the space. 
These theories appear as deformations of the familiar $AdS\times\MM$
backgrounds, and the theory is generically maximally non-local on
$\MM$ in the sense that it singles out a unique low-lying eigenvalue
of the Laplacian on $\MM$ and changes only the interactions of the
corresponding mode.

\smallskip

These theories come about as deformations of ordinary string theories
on $AdS$ spaces, via the AdS/CFT duality \refs{\ma,\witten,\gkp} (for
a review see \refs{\AharonyTI}). The spectrum of fields on the
gravitational side of the duality corresponds to a subset of the
operators in the dual CFT - these are usually referred to as ``single
trace operators''. Other operators that appear in the OPE of the
single trace operators are described by multi-particle states in the
gravitational dual. These operators are referred to as ``multi-trace
operators''. NLST's come about when there are marginal and relevant
operators that are multi-trace operators\foot{The truly marginal
deformation on $AdS$ is the main case studied so far. NLST's and
Multi-trace deformations are, however, expected to play an important
role in the relevant deformation cases as well
\refs{\evaallan}. In this paper we will discuss a marginal but not truly 
marginal deformation}. Deforming the boundary theory by such operators
does not correspond to any known stringy small shift of the background
(since these are all associated with single particle states). Hence
the theory is deformed in a new way. In \refs{\abs} it was shown that
the familiar notions of string perturbation theory themselves need
to be modified, leading to properties 1-3 above.

In this paper we will elaborate on how this deformation effects the
low energy effective action. We will show, starting from the
formulation in \refs{\abs} of the deformation using an auxiliary
variable, that the boundary conditions on some fields change. This is
a significant change in gravity, where determining the boundary
conditions is at times a subtle task, which influences physical
observables. Whereas this issue is usually neglected in smooth,
asymptotically flat spaces (where they play a small role), they are
very important in other cases such as $AdS$ and spaces with
singularities - particularly cosmological singularities. 

Motivated by this we are led to speculate what might be the more
general relation between NLST's and qualitative features of
spacetimes. We suggest that NLST may be relevant in cases where there
are severe enough singularities, defects or perhaps horizons, at
finite distance or time (or finite affine parameter) in spacetime. If
this conjecture turns out to be true, one might need to deal more
extensively with NLST's to understand such backgrounds.

In section 2 we discuss the gravity limit of the usual AdS/CFT
correspondence. Our presentation will be slightly different from the
usual presentation. This treatment is chosen because it is better
adapted to the subsequent discussion of double trace
deformations\foot{The differences are expected to disappear as the IR
cut-off in $AdS$ is removed.}. In section 3 we discuss the change in
the boundary conditions which occurs when deforming the theory by a
double trace deformation. The double trace deformation which we have
chosen is the simplest one, a deformation by $:O^2:$ where $O$ is an
operator of dimension\foot{$:O^2:$ will turn out to be marginal but
not truly marginal.} $d/2$. In section 4, which is more speculative,
we try to relate the fact that the double trace deformation changes
the boundary conditions to the non-locality of the worldsheet, and we
argue that generally when there are singularities at a finite distance
or time in spacetime, such as the boundary of AdS, the theory will be
non-local on the worldsheet.
Boundary conditions in $AdS/CFT$ correspondence where recently
discussed in \minces. In \witbou\ Witten reaches a similar conclusion regarding 
the change of 
boundary conditions in the presence of a double-trace deformation.

\newsec{$AdS/CFT$ correspondence, the usual case.}

In this section we review the familiar way in which one computes
correlators using the AdS/CFT correspondence. We will set-up the
procedure in a slightly different form than usual, that will be more
convenient when we discuss the double trace deformation. The
differences go away when we remove the IR regulators in $AdS$.

Subsection 2.1 includes the basic set-up from the literature, 2.2
discusses the modification that we need and set up notations for the
rest of the paper, and section 2.3 discusses our formulation in the
presence of sources on the boundary.

\subsec{Basic setup}

According to the AdS/CFT correspondence \refs{\ma,\witten,\gkp} (for a
review see \refs{\AharonyTI}), the generating functional of
correlation functions for some operator $\OO$ in the CFT dual to an AdS
geometry, is equal to the partition function of string theory in that
background, with specific behavior of the fields near the boundary. We
will primarily work in the Kaluza-Klein reduction down to $AdS_{d+1}$
and suppress the compact internal manifold and the 10 or 11
dimensional picture. We will also focus on the case of a single real
scalar field $\phi$ on $AdS_{d+1}$.

Imposing boundary conditions $\rho$ for $\phi$ (this will be made more
precise below, adapted for our purposes), the relation is written as
\eqn\ads{Z[\rho]_{CFT} \equiv 
<e^{\int\rho A}>_{CFT}=Z[\rho]_{string\ theory}.}
The quantity on the RHS is the full stringy partition function. Since
we are interested only in the dynamics of one scalar field we will
slightly abuse notation and write it as if it is a field theory path
integral over this scalar field. This is justified at low enough
energies before quantum gravity and stringy effects set in, which is
the regime we are interested in. Hence we will write the
correspondence as
\eqn\mgads{Z[\rho]_{CFT} \equiv <e^{\int\rho A}>_{CFT}=\int 
D[\phi;\rho]e^{-\SS_{gr}[\phi]},} where $D[\phi;\rho]$ stands for a path
integral over the field $\phi$ with boundary conditions set by $\rho$,
and $\SS_{gr}$ is the gravity action.

We work in Euclidean $AdS$ in the Poincar\'e patch with \cd s:
\eqn\metr{ds^2={dz^2+dx_idx_j\delta^{ij} \over z^2},\qquad i,j=1 \dots d,}
and we have set the length scale of $AdS$ to 1.
The action of a scalar field with mass $m$ (where the Breitenlohner-Freedman 
bound restricts $m^2 \geq -{d^2 \over 4}$) in these coordinates is:
\eqn\acgc{\SS_0=-{1 \over 2}\int 
d^dxdz\phi[\partial_z z^{-d+1}\partial_z +z^{-d+1}\partial_i\partial^i 
-z^{-d-1}m^2]\phi.}
The general solution to the equations of motion derived from \acgc\ for a 
scalar field with mass $m^2>-{d^2 \over 4}$ is 
\refs{\kw,\witten}:   
\eqn\gensol{\phi(\vec{x},z)=\int_{\partial AdS} d^dx' {\Sigma^-(\vec{x}') 
z^{\Delta_+} \over
(z^2+|\vec{x}-\vec{x}'|^2)^{\Delta_+}}+\int_{\partial AdS} d^dx'
{\Sigma^+(\vec{x}') z^{\Delta_-} \over
(z^2+|\vec{x}-\vec{x}'|^2)^{\Delta_-}},} where $\Delta_{\pm}={d \over
2} \pm \sqrt{({d \over 2})^2+m^2}$, and $\Sigma^{\pm}$ are arbitrary
functions of $x$.
Using the fact that
\eqn\deltf{\lim_{z \rightarrow 0}{z^{2\Delta_+-d} \over 
(z^2+|\vec{x}-\vec{x}'|^2)^{\Delta_+}}=\pi^{d \over2}{\Gamma(\Delta_+-{d 
\over2}) \over \Gamma(\Delta_+)}\delta^d(\vec{x}-\vec{x}'),} the asymptotic 
behavior near the boundary is 
\eqn\bobe{\phi(z,\vec{x}) \sim 
\phi_0(\vec{x})z^{\Delta_-}(1+O(z^2))+A (\vec{x})z^{\Delta_+}(1+O(z^2)),}
where $\phi_0$ and $A$ are known linear functionals of $\Sigma^-$ and 
$\Sigma^+$.

In the degenerate case $\Delta_+=\Delta_-={d \over 2}$
(i.e. $m^2=-{d^2 \over 4}$) additional regularization is required and
the second independent solution of \gensol\ has a logarithmic
term. 

\subsec{Classical solutions and boundary conditions}

In order to allow for classical solutions to \acgc, such as \gensol,
one must fix appropriate boundary conditions.  The linear variation of
\acgc\ is:
\eqn\linea{\delta \SS^{(1)}=\int_{AdS}d^{d+1}x 
\sqrt{g}\delta\phi[-\nabla^2 +m^2]\phi+{1 \over 
2}\int_{\partial (AdS)}d^dxz^{-d+1}(\phi \partial_z \delta\phi -(\partial_z 
\phi) \delta\phi).}

The first term vanishes if $\phi(\vec{x},z)$ obeys the equations of
motion in the bulk.  The second term vanishes if one fixes the
following boundary condition\foot{Note that this is not the most
general boundary condition possible. We will shortly present a more
general boundary condition.}:
\eqn\bouc{z\partial_z \phi |_{\partial}=\omega\phi |_{\partial},} where 
$\omega$ is an arbitrary function, naturally taken to be a constant, and
$\partial$ stands for $\partial (AdS)$ - the boundary of $AdS$.
This restricts the variation on the boundary to obey:
\eqn\boucv{z\partial_z \delta 
\phi|_{\partial}=\omega\delta\phi|_{\partial}.}
Plugging the general solution \gensol\ into \bouc\ where we set the
boundary at some finite cutoff\foot{We are imposing a cut off near
$z=0$ in the Euclidean Poincar\'e patch. One needs also to regulate
$z\rightarrow \infty$ since it is also a boundary point of the
geometry. This can be taken care of by requiring that all sources on
the boundary decay fast enough at $\vert \vec{x}\vert
\rightarrow\infty$.}  $z=\epsilon$ gives a complicated non-local
integral relation between $\Sigma^+$ and $\Sigma^-$ In the limit $\epsilon
\rightarrow 0$ we can use
\bobe\ instead of \gensol\ and this relation simplifies to the
following local expression:
\eqn\plug{\left[ (\Delta_--\omega)\phi_0(\vec{x})\epsilon^{\Delta_-}+ 
(\Delta_+-\omega)A (\vec{x})\epsilon^{\Delta_+} \right] 
(1+\OO(\epsilon^2))=0.} 

There are two special choices that simplify this relation
further\foot{This is different then the more familiar formulation
where one imposes Dirichlet boundary conditions on the scalar
field. The two are just different regularization schemes. The conventions we 
adopt here will be more useful below.}. If we take $\omega=\Delta_{\pm}$ we can 
write \bouc\ as:
\eqn\delm{\left[ z^{-\Delta_{\mp}}(z\partial_z-\Delta_{\pm})\phi(\vec{x},z) 
\right]_{\partial}=0,} where we multiplied \plug\ by $z^{-\Delta_\mp}$ to 
obtain a finite relation near $z\sim 0$. In the limit $\epsilon
\rightarrow 0$ this fixes the value of $\phi_0=0$ ($A=0$) but puts no
restriction on $A$ ($\phi_0$), thus recovering the usual
interpretation of $A$ ($\phi_0$) as fluctuating (in the Euclidean path
integral
\refs{\kw}) and $\phi_0$ ($A$) as a constant background. This was the
case where the classical source on the boundary is set to zero. We
will now modify the boundary condition so as to include a non-zero
source.

\subsec{Adding a classical source - finite cutoff treatment.}

The boundary condition \bouc\ makes sure that the integrand $[\phi \partial_z 
\delta \phi-(\partial_z \phi)\delta\phi]$ vanishes on the boundary.
However, we can relax this condition and demand only that the whole
integral
$\int_{\partial}d^dxz^{-d+1}(\phi\partial_z\delta\phi-(\partial_z\phi)
\delta\phi)$ will vanish on the boundary. The general boundary condition 
assuring this is:
\eqn\genbc{z\partial_z\phi(\vec{x},z)|_{\partial}=\int_{\partial}d^dx'\Omega( 
\vec{x},\vec{x}')\phi(\vec{x}',z),} where $\Omega( \vec{x},\vec{x}')$ is 
symmetric in $x$ and $x'$. 
The generalization to $\Omega$ will be important below when we discuss
the double trace deformation.  Usually one takes $\Omega=0$ for
$\vec{x}\not= {\vec{x}}'$. We will relax this here. This is reasonable
for the following two reasons:
\item{1.} We will require that the support of $\Omega$ shrinks to 
$\vec{x}={\vec{x}}'$ when we remove the cut-off from $AdS$. Hence it
is again, from the dual field theory point of view, merely a different
regularization scheme.
\item{2.} Reinstating the compact manifold that multiplies $AdS$, our 
deformation will be highly non-local in it, so we have already given up
locality to some extent anyhow.

\smallskip

For later convenience we will use infinite matrix notations, where
integrals are written as summation of indices,
\eqn\defmat{\int_{z=\epsilon} d^dx 
d^dx'\chi(\vec{x},z)F(\vec{x},\vec{x}')\psi(\vec{x}',z) \equiv 
\chi_{\vec{x}}{\bf 
F}^{\vec{x}}_{\ \vec{x}'}\psi^{\vec{x}'} \equiv \chi^T{\bf F}\psi .}
Note that the evaluation of the integral at the boundary $z=\epsilon$ is 
implicit in our notation.
The boundary condition \genbc\ will now be written as
${z\partial_z\phi|_{z=\epsilon}={\bf \Omega}\phi.}$

We can now write down the boundary condition with an external source as:
\eqn\genbcws{z^{-\Delta}(z\partial_z-{\bf \Omega})\phi|_{z=\epsilon}={\bf 
A}\rho,} where ${\bf \Omega}$ is as above and ${\bf A}$ is an
arbitrary linear functional, which encodes the renormalization of the
source with the cut-off $\epsilon$.  We will make a convenient choice
for ${\bf A}$ below, but different choices of ${\bf A}$ amount again
to different regularization schemes. \genbcws\ restricts the fluctuations
to obey $z^{-\Delta}(z\partial_z-{\bf
\Omega})\delta\phi|_{z=\epsilon}=0$ so that now the linear variation
\linea\ is:
\eqn\livre{\delta\SS^{(1)}=-{1 \over 2}\int_{z=\epsilon}d^dxd^dx'z^{\Delta-d}
\delta\phi(\vec{x},z)A(\vec{x},\vec{x}')\rho(\vec{x}')=
-{1 \over 2}z^{\Delta-d}\delta\phi^T{\bf A}\rho.}  In order to
have a classical solution we add to the action a boundary term to
cancel \livre\ giving:
\eqn\acws{\eqalign{\SS=&{1 \over 2}\int d^{d+1}x \sqrt{g}\phi[-\nabla^2 
+m^2]\phi +{1 \over 2}\int_{z=\epsilon} d^dxd^dx'
z^{\Delta-d}\phi(\vec{x},z){\bf A}(\vec{x},\vec{x}')\rho (\vec{x}')=
\cr =&\SS_0 +{1 \over 2}\epsilon^{\Delta-d}\phi^T{\bf A} \rho}}
%\SS_0+{1 \over 2}\epsilon^{-d}\phi^T(z\partial_z-{\bf \Omega})\phi.}} 

Evaluating \acws\ on a classical solution $\phi_{cl}$ one gets:
\eqn\sclold{\SS[\phi_{cl}]={1 \over 2}\epsilon^{\Delta-d}\phi_{cl}^T{\bf 
A}\rho.}
It turns out that a convenient choice reproducing the usual form of
the two point function in the $\epsilon \rightarrow 0$ limit is:
\eqn\corrc{\eqalign{{\bf \Omega}^{\vec{x}}_{\ \vec{x}'}\equiv 
&\Delta_+\delta^{\vec{x}}_{\ \vec{x}'} \cr {\bf A}^{\vec{x}}_{\
\vec{x}'}\equiv &-2\Delta_+{\epsilon^{\Delta_+-\Delta_- +2} \over
(\epsilon^2+|\vec{x}-\vec{x}'|^2)^{\Delta_+ +1}}.}}  Note, using
\deltf\ that ${\bf A}$ is proportional to a delta function as
$\epsilon\rightarrow 0$. This choice is convenient because it assures
that the solution:
\eqn\witt{\phi_{\rho}(\vec{x},z)=\int d^dx' {\rho(\vec{x}') 
z^{\Delta_+} \over (z^2+|\vec{x}-\vec{x}'|^2)^{\Delta_+}},} satisfies
the boundary condition:
\eqn\corrbc{z^{-\Delta_-}(z\partial_z-\Delta_+)\phi(\vec{x},z)|_{z=\epsilon}= 
{\bf A}^{\vec{x}}_{\ \vec{x}'}\rho^{\vec{x}'},} at any finite
$\epsilon$ by construction.

Finally, evaluating \acws\ on the classical solution \witt\ we get   
\eqn\tstbe{\SS[\rho]={1 \over 2}\int_{z=\epsilon} d^dx 
d^dx'd^dx''{\rho(\vec{x}'){\bf A}(\vec{x},\vec{x}'')\rho(\vec{x}'') \over 
(z^2+|\vec{x}-\vec{x}'|^2)^{\Delta_+}}.} This gives the desired 
result when we remove the cutoff:
\eqn\tste{\SS[\rho ]=-\pi^{d \over 2}{\Gamma(\Delta_++1-{d \over 2}) \over 
\Gamma(\Delta_+)}\int d^dxd^dx' {\rho(\vec{x})\rho(\vec{x}') \over 
|\vec{x}-\vec{x}'|^{2 \Delta_+}}.} 
Note that using this formalism the
degenerate case $\Delta_+=\Delta_-={d \over 2}$ does not pose any
problem since the exponent $\Delta$ (see \deltf) is shifted by one in
\corrc\ and becomes $\Delta +1$. The unitarity bound now emerges naturally at
$\Delta={d-2 \over 2}$ where a logarithmic divergence appears in \tstbe.

\newsec{$AdS_{d+1}/CFT_d$ deformed by a double trace operator.}

In this section we will discuss the deformation in the gravity side
caused by turning on a double trace deformation $\int d^dx {\OO}^2$ in
the dual CFT, where $\OO$ is an operator of dimension $d/2$. A ``Double
trace'' deformation is a special case of the class of ``multi trace''
deformations discussed in
\abs, and the same method described here can also be applied 
to this more general case. Also, in what follows we assume that the
gravity field $\phi$ which couples to the boundary operator is
free. This does not restrict our discussion in any important way as
interaction vertices can be easily incorporated without changing our
basic conclusions.

\subsec{The change in the boundary conditions}

As described in \refs{\abs}, One can deform the $d$-dimensional dual
CFT with ``double trace'' operators by introducing an auxiliary field
$\lambda$.
Keeping a source term for $\OO$ the expression is
\eqn\auxi{e^{{\tilde{h} \over 2}\int \OO^2+\int\rho\OO}=\int 
D[\lambda]e^{\int(\lambda+\rho)\OO-\int{1 \over
2\tilde{h}}\lambda^2}= \int 
D[\lambda]e^{\int\lambda \OO-\int{1 \over 2\tilde{h}}(\lambda-\rho)^2}, }
where we shifted variables 
$\lambda \rightarrow \lambda+\rho$ to go to the last expression.
The generating functional in the presence of the double trace operator 
deformation is therefore 
\eqn\gfcf{\eqalign{Z^{d.t.}_{CFT}[\rho] &\equiv \langle e^{{\tilde{h} \over 
2}\int \OO^2+\int\rho\OO} \rangle  =\int D[\lambda]e^{-{1 \over 
2\tilde{h}}\int(\lambda-\rho)^2}Z[\lambda]_{CFT},}}
where $Z[\lambda] \equiv \langle  e^{\int \lambda \OO} \rangle$ and the 
notation $\langle \dots \rangle$ means the path integral in the CFT 
with the undeformed action.

We now use \ads\ in the classical gravity approximation in order to move 
to the gravity side and get:
\eqn\diss{\eqalign{Z_{CFT}^{d.t.}[\rho]=&\int D[\lambda]e^{-{1 \over 2 
\tilde{h}} \int_{\partial} (\lambda-\rho)^2} Z[\lambda]_{CFT}=\int D[\lambda] 
D[\phi;\lambda]e^{-{1 \over 2 
\tilde{h}} \int (\lambda-\rho)^2}e^{-\SS_{gr}[\phi]}=Z_{gr}^{d.t.}[\rho]}.}
We will use below these expressions to define a new gravity action by
\eqn\dissb{Z_{gr}^{d.t.}[\rho]= \int D[\phi]e^{-S_{gr}[\phi;\rho;\tilde{h}]},} 
where the path integral is carried out on all $\phi$ without any boundary
conditions (this will be elaborated below).

We begin by rewriting \diss\ with the boundary condition on $\phi$ which now 
reads
\eqn\oldbc{\epsilon^{-{d \over 2}} \left((z\partial_z-{d \over 
2})\phi \right)_{z=\epsilon}={\bf A}\lambda,} imposed by a delta function
\eqn\partdb{Z_{gr}^{d.t.}[\rho]=\int D[\lambda] D[\phi]e^{-{1 \over 2 
\tilde{h}} \int (\lambda-\rho)^2}e^{-\SS_{gr}[\phi]}\prod_{\vec{x}} \delta 
\left(z^{-{d \over 2}}(z\partial_z-d/2)\phi(\vec{x},z)|_{z=\epsilon}-({\bf 
A}\lambda)(\vec{x})\right).}  The product of delta functions for each
point on the boundary can be represented by an integral on an
additional boundary field $\beta(\vec{x})$ as follows:
\eqn\partdc{Z_{gr}^{d.t.}[\rho]=\int D[\phi]D[\lambda] 
D[\beta]e^{-S[\phi,\lambda,\beta;\rho,\tilde{h}]},}
where now $\lambda,\phi,\beta$ are independent and we denoted:
\eqn\fulac{\eqalign{S&[\phi,\lambda,\beta;\rho,\tilde{h}]= \cr= &{1 \over 
2}\int_{Bulk} d^{d+1}x \sqrt{g}\phi(-\nabla^2 -{d^2 \over 4})\phi+{1 \over 
2}\int_{z=\epsilon} d^dxd^dx'z^{-d}\phi(\vec{x},z) 
(z\partial_z-{d \over2})\phi(\vec{x},z) \cr +&{1 \over 
2\tilde{h}}\int_{z=\epsilon}d^dx(\lambda-\rho)^2(\vec{x})+{i \over 
2}\int_{z=\epsilon}\beta(\vec{x}) \left[ (z\partial_z-d/2)\phi(\vec{x},z)-z^{d 
\over 2}({\bf A}\lambda)(\vec{x}) \right]=\cr =&\SS_0+{1 \over 
2}\epsilon^{-d}\phi^T(z\partial_z-{d \over 2})\phi+{1 \over 
2\tilde{h}}(\lambda-\rho)^T(\lambda-\rho)+{i \over 
2}\beta^T\left[(z\partial_z-{d \over 2})\phi-z^{d \over 2}{\bf A}\lambda 
\right].}}
We first notice that the $\lambda$ integral is Gaussian, integrating
over it we get:
\eqn\aftlamb{Z_{gr}^{d.t.}[\rho]=\int D[\phi ]D[\beta ] e^{-\SS 
[\phi,\beta;\rho,\tilde{h}]},} where
\eqn\acaflam{\SS [\phi,\beta;\rho,\tilde{h}]=\SS_0+{1 \over 
2}\epsilon^{-d}\phi^T(z\partial_z-{d \over 2})\phi+{\tilde{h} \over 
8}\epsilon^d\beta^T{\bf A}^2\beta+{i \over 2}\beta^T\left( (z\partial_z-{d 
\over 2})\phi-z^{d \over 2}{\bf A}\rho \right) .}
Now the $\beta$ integral also became a Gaussian integral so we can perform it 
and get:
\eqn\aftbet{Z_{gr}^{d.t.}[\rho]=\int D[\phi ]e^{-\SS [\phi;\rho,\tilde{h}]},}
where
\eqn\acafbet{\eqalign{&\SS [\phi;\rho,\tilde{h}]=\SS_0+\cr +&{1 \over 
2}\epsilon^{-d}\phi^T(z\partial_z-{d \over 2})\phi+{1 \over
2\tilde{h}}\epsilon^{-d} \left( (z\partial_z-{d \over 2})\phi-z^{d
\over 2}{\bf A}\rho \right)^T{\bf A}^{-2} \left( (z\partial_z-{d \over
2})\phi-z^{d \over 2}{\bf A}\rho \right).}}  Since we integrated over
the boundary condition we now have an action without any boundary
conditions. We can now ask what are the classical solutions of this
action. The linear variation is:
\eqn\linvto{\eqalign{\delta\SS=&-\int 
d^{d+1}x\sqrt{g}\delta\phi(\nabla^2+{d^2 \over 4})\phi+ \cr +&{1 \over 
\tilde{h}}\epsilon^{-d}\left( \tilde{h}{\bf A}^2\phi+(z\partial_z-{d \over 
2})\phi-\epsilon^{d \over 2}{\bf A}\rho \right)^T {\bf
A}^{-2}(z\partial_z-{d \over 2})\delta\phi.}}  The classical solutions
are now characterized by the usual equations of motion in the bulk of
$AdS$ plus a ``boundary equation of motion":
\eqn\newbc{\epsilon^{-{d \over 2}} \left((z\partial_z-{d \over 
2})\phi+\tilde{h}{\bf A}^2\phi \right)_{z=\epsilon}={\bf A}\rho.}

Equations \acafbet\ and \newbc\ are the main results of the paper. We
obtained that the path integral in the theory deformed by the double
trace deformation is a path integral without boundary conditions on
$\phi$ but with a modified boundary action, which effectively imposes
new boundary conditions on classical solutions, as can be seen by comparing 
\newbc\ with the boundary condition \oldbc.

\subsec{The induced change to the 2-point function}

We would now like to use this formulation to extract the change to
2-point functions of the operator $\OO$ on the gravity side. This will
match exactly the results expected from the field theory.

Evaluating \acafbet\ on the classical solution obeying \newbc\ we get
again (see \sclold):
\eqn\newclas{\SS [\phi_{cl};\rho,\tilde{h}]={1 \over 2}\epsilon^{-{d \over 
2}}\phi^T_{cl}{\bf A}\rho.}
We will solve the equations of motion in the bulk and on the boundary 
iteratively.
Expanding $\phi$ in powers of $\tilde{h}$ we write:
\eqn\fiwan{\phi(\vec{x},z)=\sum_{n=0}^{\infty}\tilde{h}^n\phi^{(n)}(\vec{x},z)=
\phi^{(0)}+\tilde{h}\phi^{(1)} \dots.}
Plugging this expression into \newbc\ we get:
\eqn\itter{\epsilon^{-{d \over 2}} \left((z\partial_z-{d \over 
2})+\tilde{h}{\bf A}^2 \right)(\phi^{(0)}+\tilde{h}\phi^{(1)} 
\dots)|_{z=\epsilon}={\bf A}\rho.}
The zeroth order equation is:
\eqn\ziros{\epsilon^{-{d \over 2}} (z\partial_z-{d \over 
2})\phi^{(0)}|_{z=\epsilon}={\bf A}\rho.}
The solution to that is exactly \witt.
The higher order equations are:
\eqn\hieror{\eqalign{(z\partial_z-{d \over 2}) \phi^{(1)} 
(\vec{x},z)|_{z=\epsilon}&=-{\bf A}^2\phi^{(0)}(\vec{x},\epsilon) \cr
&: \cr (z\partial_z-{d \over
2})\phi^{(n+1)}(\vec{x},z)|_{z=\epsilon}&=-{\bf
A}^2\phi^{(n)}(\vec{x},\epsilon) \cr &:}} Since ${\bf A}\rho$ does not
depend on $z$, the solution to the (n+1)'th order equation is the same
as \witt\ only with $\rho \rightarrow -\epsilon^{-{d \over 2}}{\bf
A}\phi^{(n)}(\vec{x},\epsilon).$ so in particular:
\eqn\fircht{\eqalign{\phi^{(1)}(\vec{x},z)&=-\int d^dx'\epsilon^{-{d \over 
2}}{\bf A}\phi^{(0)}(\vec{x}',\epsilon) {z^{d \over 2} \over 
(z^2+|\vec{x}-\vec{x}'|^2)^{d \over 2}}=\cr &=-\int d^dx'd^dx''d^dx'''{z^{d 
\over 2}  A(\vec{x}',\vec{x}'')\rho(\vec{x}''') \over 
(z^2+|\vec{x}-\vec{x}'|^2)^{d \over 2} (\epsilon^2+|\vec{x}''-\vec{x}'''|^2)^{d 
\over 2}}.}}
(recall that $\Delta_+=d/2$). As before, let us evaluate \newclas\ to
first order in $\tilde{h}$ by plugging in \fircht:
\eqn\fullev{\eqalign{\SS [\phi_{cl};\epsilon]=&\int d^dx_1 
d^dx_2d^dx_3{\rho(\vec{x}_1)A(\vec{x}_2,\vec{x}_3)\rho(\vec
{x}_3) \over (\epsilon^2+|\vec{x}_1-\vec{x}_2|^2)^{d \over 2}}-\cr 
-\tilde{h}&\int d^dx_1 \dots d^dx_5 
{\rho(\vec{x}_1)A(\vec{x}_2,\vec{x}_3)A(\vec{x}_4,\vec{x}_5
)\rho(\vec{x}_5) \over (\epsilon^2+|\vec{x}_1-\vec{x}_2|^2)^{d 
\over 2} (\epsilon^2+|\vec{x}_3-\vec{x}_4|^2)^{d \over 
2}}+\OO(\tilde{h}^2) .}}
When we now take $\epsilon \rightarrow 0$ and use \deltf\ we get:
 
\eqn\fullbv{\eqalign{\SS [\phi_{cl}]=&{-d\pi^{d \over 2} \over {d \over 2}!} 
\int d^dx_1 d^dx_2 {\rho(\vec{x}_1)\rho(\vec{x}_2) \over 
|\vec{x}_1-\vec{x}_2|^d }-\cr -\tilde{h}\left({-d\pi^{d \over 2} \over
{d \over 2}!} \right)^2 &\int d^dx_1 d^dx_2 d^dx_3
{\rho(\vec{x}_1)\rho(\vec{x}_3) \over |\vec{x}_1-\vec{x}_2|^d
|\vec{x}_2-\vec{x}_3|^d}+\OO(\tilde{h}^2) .}}  This is the term that
upon varying twice with respect to $\rho$ gives the right leading
$\tilde{h}$ correction to the two point function in the CFT in the limit 
$\epsilon \rightarrow 0.$ 
\eqn\cftexp{\eqalign{<\OO(\vec{x})\OO(\vec{y})>={1 \over 
|\vec{x}-\vec{y}|^d}+&\tilde{h}\int d^du {1 \over|\vec{x}-\vec{u}|^d 
|\vec{u}-\vec{y}|^d}+ \cr +&\tilde{h}^2 \int d^dud^dv {1 
\over|\vec{x}-\vec{u}|^d |\vec{u}-\vec{v}|^d |\vec{v}-\vec{y}|^d}+ \dots,}}
as in e.g. equation (2.4) in \abs.  It is also clear that the higher order
corrections will follow the same pattern, where in each order one adds
another boundary integration and another boundary-to-boundary
propagator.

It is worth revisiting one point. \newbc\ is an equation of motion on
the boundary and not a boundary condition like
\genbcws.  The important difference between these two is that boundary
conditions restrict the quantum fluctuations one is integrating over
while an equation of motion on the boundary does not. One might ask
whether we should also impose \newbc\ as a boundary condition in the
quantum path integral. At the level that we analyzed the 2-point
functions we can not answer this question since we used only the
classical configurations. Only when we start considering quantum loops
in $AdS$ will the difference between the two formulations appear.

\subsec{Renormalization - the CFT side.}

The expansion of the two point function in the CFT deformed by the
double trace operator is of the form \cftexp.  Notice that already the
first correction diverges when the intermediate coordinate $\vec{u}$
approaches either $\vec{x}$ or $\vec{y}$, and the following orders
diverge correspondingly, so one needs to regularize these expressions.
This can be done by renormalizing the operator $\OO$ as follows:
{}First one introduces an ultraviolet cutoff $\epsilon$ which prevents
$\vec{u}$ from approaching either $\vec{x}$ or $\vec{y}$ too closely,
which can be done by changing the propagator to:
  
\eqn\cfteps{<\OO(\vec{x})\OO(\vec{y})>={1 \over 
(\epsilon^2+|\vec{x}-\vec{y}|^2)^{d \over 2}}+\tilde{h}\int {d^du
\over (\epsilon^2+|\vec{x}-\vec{u}|^2)^{d \over 2}
(\epsilon^2+|\vec{u}-\vec{y}|^2)^{d \over 2}}+\dots,} where the
specific form we chose for the cutoff is suggested by the $AdS$ dual
but in the end of the day we will anyways take it to zero (and 
be left with cutoff independent answers).

It is easy to identify the divergent part of the integral:
\eqn\fulesp{\eqalign{\tilde{h}\int &{d^du \over 
(\epsilon^2+|\vec{x}-\vec{u}|^2)^{d \over 2}
(\epsilon^2+|\vec{u}-\vec{y}|^2)^{d
\over 2}}=\cr &={\tilde{h} \over |\vec{x}-\vec{y}|^d} 
\left( 2\Omega_dlog({|\vec{x}-\vec{y}|\over \epsilon}) +finite +O({\epsilon
\over |\vec{x}-\vec{y}|}) \right),}}where $\Omega_d$ 
is the volume of the $d-1$ unit sphere. 

This phenomenon is familiar in field theory and is usually dealt with by
renormalizing the operator $\OO$ order by order in conformal
perturbation theory \refs{\DijkgraafJT}. We thus define a renormalized operator
$\widehat{\OO}$ as follows:
\eqn\redoo{\widehat{\OO}(\vec{x}) \equiv \OO(\vec{x})\left(1-\tilde{h} 
\Omega_d log({\Lambda \over \epsilon})\right),} where $\Lambda$ is an arbitrary 
scale needed because we are renormalizing $\OO(\vec{x})$ which can not depend on 
$\vec{y}$.

Clearly in the two point function of $\widehat{\OO}$ no divergences
appear to first order in $\tilde{h}$ and one gets:
\eqn\tpforo{\eqalign{<\widehat{\OO}(\vec{x})\widehat{\OO}(\vec{y})>  &= {1 
\over |\vec{x}-\vec{y}|^d}-\tilde{h} { 2\Omega_dlog({\Lambda 
\over \epsilon}) \over |\vec{x}-\vec{y}|^d}+\tilde{h}\int {d^du \over 
(\epsilon^2+|\vec{x}-\vec{u}|^2)^{d \over 2}
(\epsilon^2+|\vec{u}-\vec{y}|^2)^{d \over 2}}+ O(\tilde{h}^2) \cr &={1
\over |\vec{x}-\vec{y}|^d}+\tilde{h} {2\Omega_dlog({|\vec{x}-\vec{y}|
\over \Lambda}) \over |\vec{x}-\vec{y}|^d}+\tilde{h}{finite \over
|\vec{x}-\vec{y}|^d} +O(\tilde{h}^2) .}}Note that to first
order in $\tilde{h}$ this looks like a correction to the dimension of
$\OO$. However the perturbation actually breaks conformal invariance
as can be seen by the following simple argument:\foot{We thank
D.Kutasov for pointing this out for us.}  The same calculation can now
be performed to compute corrections to the two point function of the
double trace deformation itself,
i.e. $<\OO^2(\vec{x})\OO^2(\vec{y})>$, showing that the same kind of
logarithmic divergences appear, breaking conformal invariance. Alternatively one 
can use the Callan-Symanzik
equation to see that the beta function is no longer zero.

\subsec{Renormalization - the gravity side.}

The divergences we encountered on the CFT side appear also in the
 gravity side. Indeed if we want to have a sensible description of the
 deformations caused by the insertion of the double trace operator, we
 must see that the deformed field configuration \fiwan\ is
 finite. Since the corrections to $\phi$ produce via \newclas\ the
 corrections to the gravity action (which in turn change the two point
 function) one can easily see that the correction to first order in
 $\tilde{h}$ \fircht\ diverges in exactly the same way as
 \fulesp. The reason for this divergence is that the
 ``source'' of $\phi^{(1)}$ which is $\epsilon^{-{d \over 2}}{\bf
 A}\phi^{(0)}(\vec{x},\epsilon)$ diverges.  We now show how to
 regularize the gravity action so as to get a finite answer for
 \fircht\ and then show that this regularization could have been
 easily ``guessed'' from the CFT regularization in the previous
 section.  The way we chose to regularize the gravity theory is by
 changing the boundary condition \newbc\ to:
\eqn\regbc{\epsilon^{-{d \over 2}} \left((z\partial_z-{d \over 
2})\phi+\tilde{h}{\bf A}^2\phi \right)={\bf
A}\left(1-\tilde{h}\Omega_dlog({\Lambda \over
\epsilon})+O(\tilde{h}^2) \right)\rho.}  The zeroth order equation
remains the same as \ziros\ and we rewrite \witt\ in matrix notations
for brevity as $\phi^{(0)}(\vec{x},z)={\bf B}^z\rho$.  The first order
equation changes upon regularization to:
\eqn\firreg{\eqalign{&\epsilon^{-{d \over2}}(z\partial_z-{d \over 2}) \phi^{(1)} 
(\vec{x},z)|_{z=\epsilon}=-{\bf A}\left( {\bf A}\epsilon^{-{d
\over2}}\phi^{(0)}(\vec{x},\epsilon)+\Omega_d Log({\Lambda
\over \epsilon}) \rho(\vec{x})\right)=\cr = &-{\bf A}\left( {\bf
A}\epsilon^{-{d \over2}}{\bf B}^{\epsilon}+\Omega_d
Log({\Lambda \over \epsilon})\right) \rho.}}  This way we simply
removed from the source of $ \phi^{(1)}$ the divergent part, making it finite, 
and when computing now the two point function we will get
the correct finite expression \tpforo\ to first order in $\tilde{h}$.
One could have ``guessed'' this correction by the following argument:
The source term for $\OO$ in the CFT is $\int \rho \OO$. Since the
``physical'' operator is the renormalized one \redoo\ the source term
should be changed to:
\eqn\newsou{\int d^dx \rho(\vec{x}) \widehat{\OO}(\vec{x})=\int d^dx 
\rho(\vec{x})\left(1-\tilde{h} \Omega_d log({\Lambda \over 
\epsilon})+O(\tilde{h}^2) \right)\OO(\vec{x}),} thus following the same 
calculations done in section 2 we will get the new form of the boundary 
condition \regbc.
It is easy to get convinced that this procedure can be continued to
all orders in $\tilde{h}$ in both sides of the correspondence
simultaneously.
 
\newsec{Spacetime singularities and non-local worldsheets}

{}\foot{Parts
 of this section were developed in collaboration with E. Silverstein}
We have seen in the previous section that the effect of the double
trace deformation is to change the boundary conditions on the fields
in $AdS$. Given that determining boundary conditions (and boundary
terms) in General Relativity, especially at singularities, is a subtle
yet crucial step in any actual computation, this is a significant
change.

The modification to the worldsheet formulation \refs{\abs} is even
more dramatic - the worldsheet field theory becomes explicitly
non-local, with an action of the form
\eqn\wrldshta{
S=\int d^2 z {\cal L}_0 + {\tilde h} \Sigma_{I,J}c_{IJ}\int\int d^2z_1
d^2z_2 V^{(I)}(z_1,{\bar z}_1) V^{(J)}(z_2, {\bar z}_2), } where
${\cal L}_0$ is a local functional of the worldsheet fields, the $V$'s
are vertex operators on the worldsheet, and $c_{IJ}$ are specific numbers. The
2nd term on the RHS of \wrldshta\ is an explicitly non-local
contribution to the action, which directly couples the fields at two
separate points $(z_1,{\bar z}_1)$ and $(z_2,\bar z_2)$ on the
worldsheet.

In this section we will attempt to speculate on the relation of these
two facts. We will suggest that
\item{1.} The phenomena of non-local worldsheet theories might occur 
in a much wider class of stringy backgrounds, 
\item{2.} In many cases the non-localities
would be of an even more severe type than \wrldshta, 
\item{3.} This behavior will be associated with singularities or boundaries 
in spacetime
(we of course understand many kinds of singularities in String theory,
but hardly all).
\smallskip
\noindent Unfortunately since we do not yet have a concrete workable 
example\foot{In a ``work in progress'' project we are trying to
develop such examples}, this section is highly speculative.

This might put in a broader context the results of the previous
section that the deformation effects the boundary conditions. If the
non-local deformation of the worldsheet is associated with
submanifolds of spacetime (boundaries or singularities) it can only
manifest there at low energies as boundary conditions\foot{This,
however, immediately implies effects in the bulk, away from the
boundary/singularity, as processes can communicate with the
boundary/singularity. This should perhaps be viewed as ``finite
volume/distance effects''. Reference
\refs{\nlst} clearly exhibits such effects.}.

More precisely, we will suggest that the boundaries/singularities
associated with a non-local worldsheet are ones at finite distance in
spacetime. $AdS$ certainly falls into that category as particles with
appropriate masses can propagate to the boundary of spacetime and come
back at finite time - they view the boundary as a finite distance
defect from which they scatter.

To justify these conjectures we begin by discussing whether there are
any reasons why the worldsheet has to be local. After all, it is an
auxiliary object with, for the most part\foot{with notable
exceptions, such as the interpretation of macroscopic strings in
AdS/CFT as Wilson lines \refs{\MaldacenaIM, \ReyIK}.}, only indirect
consequences for space time. One answer is that it works remarkable
well, and might therefore be treated as an axiom. A more substantive
answer is that taking the worldsheet to be a local theory is
``technically natural'' for most of the backgrounds studied so far in
string theory (which are of the form $flat\ space
\times\ compact\ manifold$). This naturality argument, however, may not prohibit 
non-local worldsheets for other more complicated backgrounds (such as
$AdS$ where we see a departure from worldsheet locality).

By worldsheet locality being ``technically natural'' we refer to the
Fischler-Susskind mechanism \refs{\FischlerTB, \FischlerCI}. For
completeness we will briefly review this mechanism: consider string
theory on some tachyon free (at tree level) non-supersymmetric
background with a 1-loop induced dilaton tadpole. Other computations
on the torus, and certainly at higher genus, would be divergent
because of this tadpole. An example of this, given in
\refs{\FischlerTB}, is of an n-point function of vertex operators 
on a torus. The divergence comes from the boundary of the moduli space
of the punctured torus where it looks like a sphere with an
infinitesimal handle, well separated from the vertex operators.  The
diagram then factorizes into a tadpole of the dilaton at zero momentum
on an infinitesimal torus, an n+1-point function on the sphere involving
the n operators we had before + the dilaton vertex operator, and a
dilaton propagator between these 2 components. It is then shown in
\refs{\FischlerTB,\FischlerCI} that this region leads to a
divergence in the n-point function. 

Fortunately one can ``renormalize'' this divergence, at the price of
shifting the background to that of a cosmological constant - this is
the Fischler-Susskind mechanism. This is done by adding a divergent
counterterm to the worldsheet theory already at the sphere level. The
main point is that because this counterterm originates from a torus
which looks like an infinitesimal handle attached to a sphere, it
yields a local correction to the worldsheet action. Hence the
worldsheet locality is preserved under quantum renormalization of the
theory - it is ``technically natural''.

We can now discuss when a violation of worldsheet locality might be
forced upon us. An important point in the discussion above was that
the divergence was associated with an on-shell dilaton tadpole. This
was used for example when factorizing the diagram to separate the
small handle from the vertex operators in the diagrams, which was an
essential step in renormalizing the theory. A particle going on-shell
implies an that it can propagate for a long time in spacetime. Hence
we expect not to be able to factor out the small handles when the
divergences are associated with a singularity or a defect in spacetime
which particles can reach at {\it finite time}. In this case, it might
be that one will encounter a divergence on the torus, but one will not
be able to isolate the singularity, as we did before, as associated
with a distinct very small handle. Rather the divergence will be
associated with a torus at a finite modular parameter. If the theory
is to make sense, one needs to absorb this divergences on the
sphere. Even if this is possible, it will be done not by an
infinitesimal handle attached to a sphere, but rather by a macroscopic
handle. This will make the worldsheet theory on the sphere non-local.

It seems that one can therefore suggest that divergences associated
with finite distance/time processes in spacetime will be reflected in
string theory by having a non-local worldsheet. We do not however have
a concrete example of this type\foot{It seems that one might be able
to build models that have this property. They are associated with
non-trivial singular cosmologies. Other relations of non-local
worldsheets to cosmology are being studied in \refs{\futwork}.}, hence
it is difficult to really check this conjecture. We will therefore
turn our attention to another speculative example, that of string theory 
on Rindler space\foot{Parts of this analysis were developed jointly with
E. Silverstein. This example is intimately related to that of particle
creation in cosmological setting which will be elaborated in
\refs{\futwork}.}, which although far from being understood, 
points to the same direction.
Let us adopt the coordinate system in
which the Rindler metric is static (we follow the conventions of
\refs{\bd})
\eqn\rndlr{
ds^2=e^{2z}(-dt^2+dz^2).  }  The Rindler horizon is at $z\rightarrow
-\infty$. As a first guess one might try and write down the string
worldsheet action using the metric \rndlr. However, since the
worldsheet theory is target-space reparametrization invariant, the
action is the same as part of flat space, and there is no suppression
in the worldsheet path integral for a string trying to cross the
Rindler Horizon. However, we do not want to allow the string to do
so because
\item{1.} the accelerating Rindler observer can not receive
information from behind the horizon.
\item{2.} If we re-enstate the entire spacetime
we will be forced to the ordinary Minkowski vertex operators and
Minkowski vacuum, and will end up with ordinary Minkowski space.

\noindent Hence it seems that we have reached an impasse.

To try and make progress we note that the problem has a similar flavor
to the one we have been discussing - there is a boundary (more
generally - some defect) in spacetime which the particles can reach at
finite proper distance, and where one needs to define the boundary
conditions (this is one of the differences between a quarter of
Minkowski space and Rindler space). Let us therefore try to apply the
ideas mentioned above about non-local worldsheets. Again
unfortunately, the picture that we will present will be very
incomplete but, we believe, suggestive and in line with the discussion
above of worldsheet non-localities.

We want to reach a point where we discuss only the part of the
worldsheet outside the horizon, and measurement processes outside the
horizon. As explained above, however, the worldsheet can easily cross
the horizon without any suppression in the action. We will therefore
organize the perturbation theory in the following way. We will split
the space of the configurations of the worldsheet - the measure in the
worldsheet path integral - into different regions depending on how
many times the worldsheet crosses the horizon. Hence a sphere which
has $k$ prongs going into the horizon will be part of the region
associated with a sphere with $k$ holes (see figure 1). Similarly a
torus which goes from one side of the horizon to the other looks to
the outside observer as a sphere with 2 holes (figure 2). After
splitting the diagram over what the outside observer sees, one
performs the path integral over the region behind the horizon (we will
not require the details at the level of the discussion here).

\fig{A sphere diagram in Rindler space with k prongs reaching beyond the 
horizon}{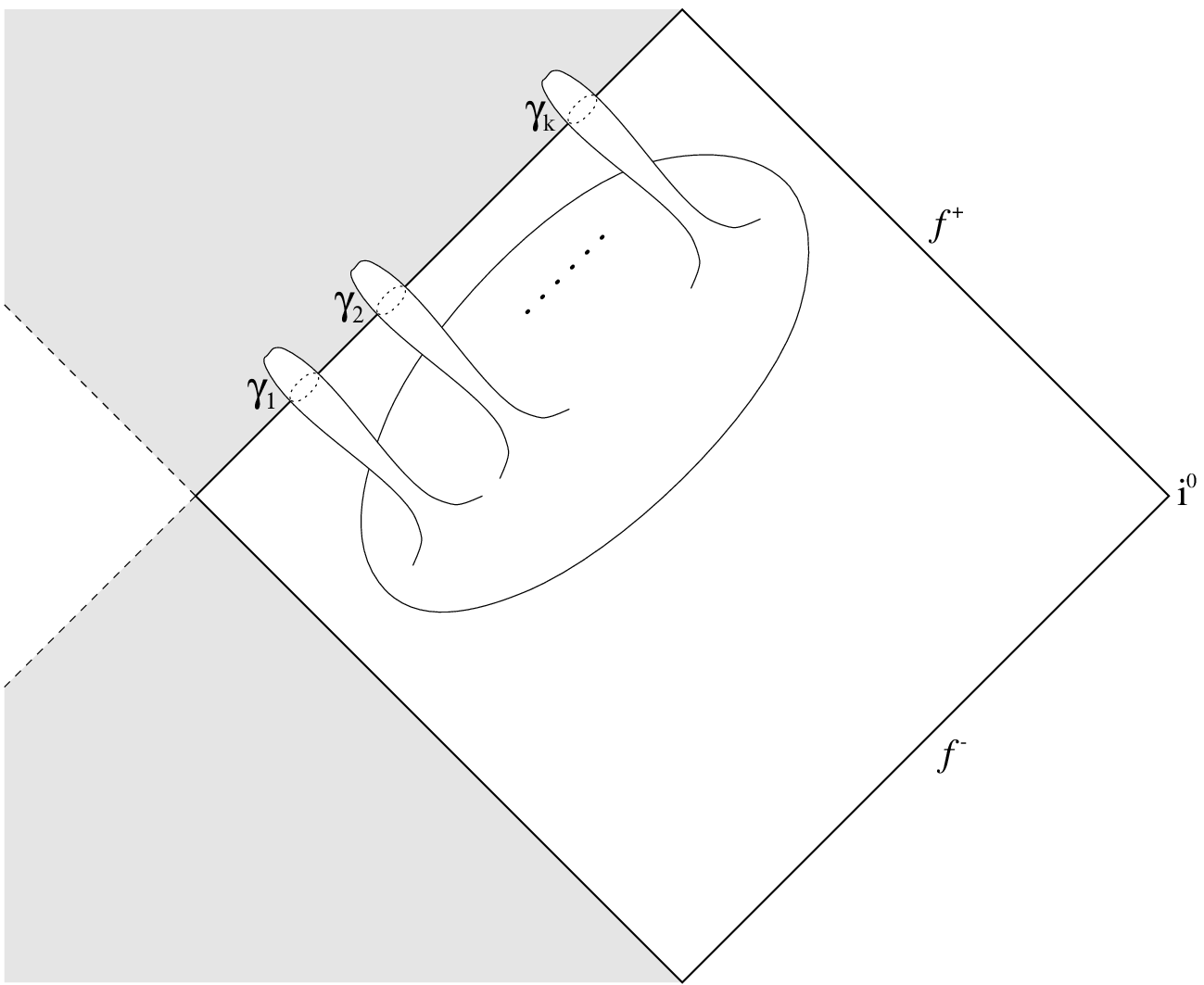}{6 truecm}
\figlabel\sphrhrzn

\fig{A torus diagram in Rindler space which crosses the horizon along 2 
circles}{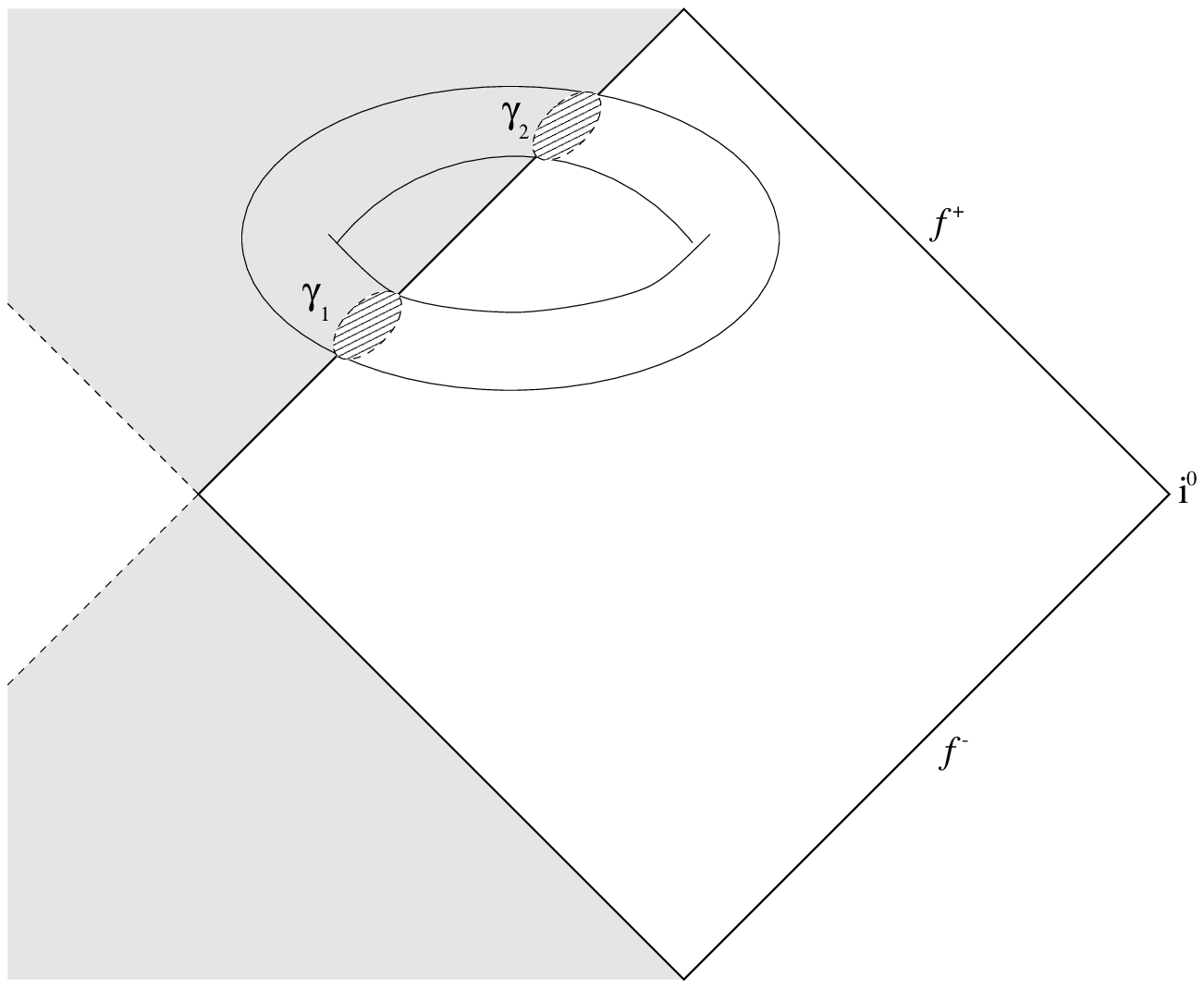}{6 truecm}
\figlabel\trshrzn

When we have a boundary where the string worldsheet crosses the
horizon (shaped like a circle on the worldsheet) we need to specify
the boundary conditions there. The standard way of doing so is by
specifying the closed string state along each of the circles. Let us
first consider the diagram in figure 1. Since each of the prongs that
goes behind the horizon is independent from the others, the states on
the different boundary circles are not correlated. The path integral
is therefore carried out with some ``boundaries multi-state''
\eqn\bndrya{
\Pi_i \bigl(\Sigma_{(I_i)} C^{(I_i)}_i\vert \phi^{(I_i)} >_{\gamma_i}\bigr) 
} where $\vert \phi^{(I)} >$ run on all the states in the closed
string channel as the index $I$ varies, and $C^{(I)}_i$ are some
numbers\foot{In the case where there is only a simple cap behind the
horizon, only states in the conformal block of the identity
contribute.}. This is still to be considered a local deformation of
the worldsheet since the states in different points of the worldsheets
- the different circles - are uncorrelated.

The situation is different in figure 2 (torus crossing the horizon) -
since the 2 circles are connected behind the horizon, the 2 states on
the 2 boundary circles will be correlated. One therefore inserts
the following ``boundaries multi-state'':
\eqn\bndryb{
\Sigma_{(I,J)} C^{(I,J)}     \vert \phi^{(I)} >_{\gamma_1} 
			\times \vert \phi^{(J)} >_{\gamma_2} } The
theory has now become non-local on the worldsheet since the behavior at
different points is correlated explicitly by hand. Hence we see
another example where a defect, in this case an horizon, at finite
distance in spacetime, is potentially giving rise to a non-local
worldsheet action.

One can easily see that the arguments leading to equation \bndryb\ can
be applied to any singularity which all the particles can reach
without being forced on shell. This is encoded by the fact that all
the single closed string states can appear as factors in the sum, or
correspondingly that one has an entire circle (which can
carry the information about the various states). This is actually a
more pathological behavior than we have seen in the $AdS$ case. One
can pass, however, from this to the $AdS$ as a special case. An
important feature of the boundary of $AdS$ is that only a finite
number of particles can approach it in finite time. Hence the sum over
the states \bndryb\ collapses into a sum over a finite number of
states. In addition, the worldsheets in AdS are pinched to a point as
they reach the boundary \refs{\deBoerPP}. The combination of these two
points collapses the more general deformation \bndryb\ into a sum over
a finite number of local vertex operators at different points on the
worldsheet, i.e, of the form \wrldshta.

\vskip 1cm

\centerline{\bf Acknowledgments}

We would like to thank T. Banks, D. Berman, S. Elitzur, A. Giveon,
D. Kutasov, J. Maldacena, E. Martinec, E. Rabinovici, E. Silverstein
and L. Susskind for helpful discussions. We are particularly indebted
to O. Aharony for collaboration at early stages of this work. M.B. is
supported by the Israel-U.S. Binational Science Foundation, the IRF
Centers of Excellence program, the European RTN network
HPRN-CT-2000-00122, and by the Minerva foundation. A.Sh is supported
by a Clore fellowship.

\listrefs

\end